\documentclass[a4paper,conference]{IEEEtran}
\IEEEoverridecommandlockouts

\usepackage{algorithmic}
\usepackage{graphicx}
\usepackage{textcomp}
\usepackage{xcolor}
\usepackage{url}
\usepackage{tabularx}
\usepackage{adjustbox}
\usepackage{multirow}
\usepackage{booktabs}
\usepackage{colortbl}
\usepackage{threeparttable}
\usepackage{cite}
\usepackage{amsmath}
\usepackage{amssymb}
\usepackage{siunitx}

\def\BibTeX{{\rm B\kern-.05em{\sc i\kern-.025em b}\kern-.08em
    T\kern-.1667em\lower.7ex\hbox{E}\kern-.125emX}}

\begin{document}

\title{Mamba-Family State-Space Model Kernels \\on a Programmable CGLA}

\author{
\IEEEauthorblockN{Takuto Ando, Yasuhiko Nakashima}
\IEEEauthorblockA{Nara Institute of Science and Technology, 8916-5 Takayama-cho, Ikoma, Nara 630-0192, Japan\\
\texttt{ando.takuto.an5@naist.ac.jp}}
}

\maketitle

\begin{abstract}
Edge and embedded inference is constrained by power and data movement.
Mamba-family state-space models replace attention with sequence-linear recurrence, but their inference path combines dense projections, short-reduction SSD kernels, and recurrent-state updates.
This paper maps these kernel groups onto IMAX, a programmable CPU-Grounded Linear Array (CGLA), and measures them from kernel execution to token-level integration.
Projection kernels match the long-reduction IMAX pipeline, whereas SSD Step-1 is limited by short reductions and kernel-boundary overheads.
Mamba-130M token-level integration identifies projection GEMV as the decode bottleneck.
These results show that programmable CGLAs fit long-reduction projection kernels, while SSD and decode-time projection support require boundary reduction and persistent-weight execution.

\end{abstract}


\bstctlcite{ssmBSTcontrol}

\section{Introduction}
\label{intro}

Edge deployment of sequence models is constrained by inference power and data movement.
Attention-based models are costly for long contexts on edge systems.
State-space models (SSMs), including Mamba and Mamba-2~\cite{mamba1,mamba2}, replace attention with sequence-linear recurrence.

The hardware question is not whether an SSM layer is one accelerator kernel.
A Mamba-2 layer mixes long dense projections, short-reduction SSD kernels, and sequential recurrent-state updates.
This mix stresses a programmable CGLA without requiring a fixed-function Mamba datapath.
Although these kernels are mostly MAC-dominated, they differ in reduction length, pipeline occupancy, data movement, and recurrence dependency.

We implement and measure the main Mamba-family kernel groups on IMAX~\cite{imax_access}.
IMAX is a programmable CPU-Grounded Linear Array (CGLA) that couples a host CPU to a one-dimensional streaming pipeline of processing elements (PEs) and Local Memory Modules (LMMs).
The same execution stack is used for projection, SSD Step-1, recurrent-state update, and token generation.
The measurements report IMAX execution time, DMA-inclusive total time, an ARM Cortex-A72 software reference, and token-integration behavior.
The focus is kernel suitability and bottleneck diagnosis rather than peak GPU competition.

The main contributions of this paper are as follows:
\begin{itemize}
  \item We implement projection, SSD Step-1, Hadamard, and recurrent-state kernels on IMAX and measure them under a common profiling path.
  \item We characterize which programmable-CGLA paths match long-reduction, short-reduction, and recurrent Mamba kernels.
  \item We integrate an IMAX offload path in Mamba-130M autoregressive decoding and identify projection GEMV, rather than the recurrent state update, as the dominant decode bottleneck on the current programmable CGLA stack.
\end{itemize}

The remainder of this paper is organized as follows.
Section~\ref{rwork} reviews related work in Mamba hardware and programmable dataflows.
Section~\ref{background} introduces Mamba-370M, IMAX, and the measurement scope.
Section~\ref{proposed} describes the IMAX-Mamba mapping and the kernel-characterization setup.
Section~\ref{ex_and_re} reports performance against the ARM software reference and separates measured FPGA timing from ASIC model context.
Section~\ref{conclusion} concludes the paper and outlines future work.

\section{Related Work}
\label{rwork}

\begin{table*}[t]
  \centering
  \caption{Position of this work relative to Mamba/SSM accelerators.
           This paper is a programmable-CGLA diagnostic study rather than a dedicated Mamba accelerator design.}
  \label{tab:related_taxonomy}
  \footnotesize
  \renewcommand{\arraystretch}{1.12}
  \setlength{\tabcolsep}{3pt}
  \begin{tabularx}{\textwidth}{@{} >{\raggedright\arraybackslash}p{2.1cm} >{\raggedright\arraybackslash}p{2.1cm} >{\raggedright\arraybackslash}X >{\raggedright\arraybackslash}X @{}}
    \toprule
    \textbf{Work} & \textbf{Datapath} & \textbf{Primary strategy} & \textbf{Contrast to this work} \\
    \midrule
    LightMamba~\cite{lightmamba} & FPGA dataflow & Quantization and hardware co-design & Optimizes a Mamba-specific deployment path. \\
    FastMamba~\cite{fastmamba} & FPGA dataflow & Quantization and approximate nonlinear units & Optimizes a Mamba-specific deployment path. \\
    MARCA~\cite{marca} & ASIC & Mamba operator specialization & Adds architecture support for the SSM operator. \\
    MARCA-v2~\cite{marcav2} & ASIC & Sparse element-wise support for SSMs & Adds metadata and sparsity support for Mamba operators. \\
    MambaOPU~\cite{mambaopu} & FPGA overlay & SSD operator fusion and sparse skipping & Builds an SSD-specific overlay path. \\
    SpecMamba~\cite{specmamba} & FPGA & Speculative decoding for Mamba inference & Changes the autoregressive execution strategy. \\
    EpochCore~\cite{epochcore} & Systolic + rec.\ PE & Dedicated recurrence support with GEMM & Adds recurrent-state processing elements. \\
    Our IMAX prior work~\cite{imax_access,ieeeaccess,whisper} & Programmable CGLA & Linear pipeline for GEMM, LLM, and ASR workloads & Does not evaluate Mamba kernels. \\
    \midrule
    \textbf{This work} & Programmable CGLA & Diagnostic implementation without fixed-function SSM datapaths & Separates projection, SSD, recurrence, and decode bottlenecks on one stack. \\
    \bottomrule
  \end{tabularx}
\end{table*}

Dedicated accelerators such as TPU~\cite{TPU}, DianNao~\cite{diannao}, and Eyeriss~\cite{eyeriss} specialize in matrix-dominated inference workloads.
Coarse-grained reconfigurable architectures pursue programmable spatial fabrics for loop kernels~\cite{cgra_survey,cgra1,cgra_cnn2}.
IMAX belongs to the programmable spatial-fabric line.
Our prior IMAX studies evaluate GEMM, LLM, and Whisper workloads on a programmable CGLA~\cite{imax_access,ieeeaccess,whisper}.
Table~\ref{tab:related_taxonomy} summarizes how the Mamba/SSM accelerators differ from the programmable-CGLA diagnostic scope of this work.

Mamba-focused hardware studies include LightMamba~\cite{lightmamba} and FastMamba~\cite{fastmamba} for quantized FPGA dataflow.
MARCA~\cite{marca} and MARCA-v2~\cite{marcav2} target Mamba operator specialization.
MambaOPU~\cite{mambaopu} targets SSD with an FPGA overlay and operator fusion.
SpecMamba~\cite{specmamba} targets SSM-specific speculative decoding.
Each study modifies the model, the datapath, or the autoregressive execution strategy.
EpochCore~\cite{epochcore} adds dedicated processing elements for the recurrent integration alongside a systolic GEMM array.
Fine-grained fusion studies also identify operator fusion and scheduling as central issues for SSM acceleration~\cite{finegrainedfusion}.
These designs treat Mamba inference as more than a GEMM problem.
Long reductions, state recurrences, element-wise operations, and memory traffic expose different datapath needs.

Unlike Mamba-specific FPGA and ASIC accelerators, this work treats Mamba inference as a diagnostic workload for an existing programmable CGLA stack.
It does so without model changes or fixed-function SSM units.
The goal is to identify which Mamba phases fit the CGLA pipeline and which phases are limited by short reductions, kernel boundaries, or token-level projection GEMV.

\section{Background}
\label{background}

\subsection{Mamba-2 Layer with Three Kernel Shapes}

The Mamba-2 model~\cite{mamba2} reformulates selective SSMs as a Structured State Space Duality (SSD).
A single layer has model dimension $d_\text{model}$, inner dimension $d_\text{inner} = 2 d_\text{model}$, $H$ heads of size $d_\text{head}$, state size $N$, and chunk size $Q$.
It proceeds in the phases summarized in Table~\ref{tab:phase_shape}.

\begin{table}[t]
  \centering
  \caption{Mamba-2 phases on Mamba-370M ($d_\text{model}{=}1{,}024$, $d_\text{inner}{=}2{,}048$, $H{=}32$, $d_\text{head}{=}64$, $N{=}128$, $Q{=}64$).
           The table lists kernel shape, reduction length, and pipeline-fit prediction for a 64-stage IMAX lane.}
  \label{tab:phase_shape}
  \footnotesize
  \renewcommand{\arraystretch}{1.15}
  \setlength{\tabcolsep}{3pt}
  \begin{tabularx}{\columnwidth}{@{} l l c >{\raggedright\arraybackslash}X @{}}
    \toprule
    \textbf{Phase} & \textbf{Kernel} & \textbf{Red.\ len.} & \textbf{64-stage fit} \\
    \midrule
    in-proj    & dense GEMM      & $L{=}1{,}024$ & long, deep occupancy \\
    out-proj   & dense GEMM      & $L{=}2{,}048$ & long, deep occupancy \\
    SSD Step-1 & GEMM1+GEMM2     & $N,Q{\leq}128$ & short, setup-cost dominated \\
    Step-3     & MV recurrence   & $N$ (sequential) & sequential dependency \\
    \bottomrule
  \end{tabularx}
\end{table}

The projection phases are the dense in-proj and out-proj GEMMs.
The input $u \in \mathbb{R}^{T \times d_\text{model}}$ is linearly projected to the concatenated features $[z, x, B, C, \Delta]$.
\begin{equation}
  [z, x, B, C, \Delta] = u\, W_\text{in},
  \label{eq:inproj}
\end{equation}
where $W_\text{in} \in \mathbb{R}^{d_\text{model} \times d_\text{in}}$ and $d_\text{in} = 2 d_\text{inner} + 2 N + H$ for standard SSD with $n_\mathrm{groups}=1$.
For Mamba-370M, the inner dimension of this GEMM is $L = d_\text{model} = 1{,}024$.
The output projection is a second dense GEMM.
\begin{equation}
  y = (o \odot \sigma(z))\, W_\text{out},
  \label{eq:outproj}
\end{equation}
with $W_\text{out} \in \mathbb{R}^{d_\text{inner} \times d_\text{model}}$ and inner dimension $L = d_\text{inner} = 2{,}048$ for Mamba-370M.
Equations~\eqref{eq:inproj} and~\eqref{eq:outproj} are the dense projection kernels mapped to the IMAX GEMM path in Section~\ref{proposed}.

SSD Step-1 handles intra-chunk computation after the sequence is divided into chunks of length $Q$.
\begin{align}
  M &= (C \cdot B^\top) \odot \Lambda \quad &&\text{(GEMM1 + Hadamard)},
    \label{eq:ssd_step1}\\
  Y &= M \cdot X                       \quad &&\text{(GEMM2)}.
\end{align}
In Eq.~\eqref{eq:ssd_step1}, $C$ and $B$ denote $Q \times N$ state factors, $\Lambda$ denotes a $Q \times Q$ lower-triangular mask, $X$ denotes the $Q \times d_\text{head}$ input block, and $Y$ denotes the $Q \times d_\text{head}$ output block.
Here, $\odot$ denotes element-wise multiplication.
GEMM1 has reduction length $N$, while GEMM2 has reduction length $Q$.
Both reductions are short relative to the 64-stage IMAX FMA chain.
The measured Step-1 sweep uses these short reductions, and Section~\ref{ex_and_re} reports the representative $N{=}Q{=}64$ kernel configuration separately from the full-layer FLOP.

Step-3 updates the recurrent state $h \in \mathbb{R}^{N \times d_\text{head}}$ sequentially across chunks.
\begin{equation}
  h_c = A_c h_{c-1} + B_c^\top X_c,
  \label{eq:step3}
\end{equation}
where $A_c$ is an input-dependent diagonal matrix.
The chunk dependency prevents parallel execution across chunks.
This phase is therefore sequential.

\begin{figure*}[!t]
  \centering
  \includegraphics[width=\textwidth]{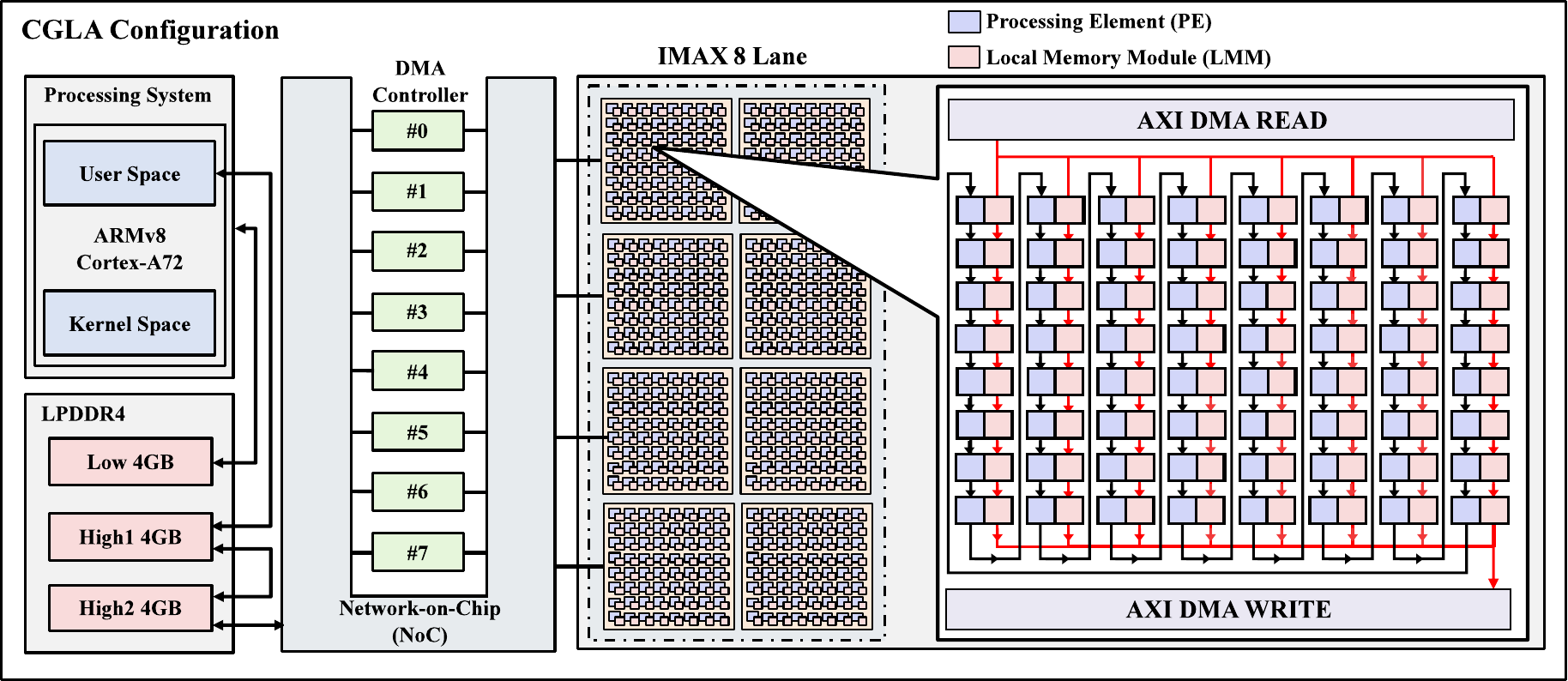}
  \caption{IMAX3 FPGA prototype.
           The PL supports an eight-lane IMAX configuration, while the measurements in this paper use one CGLA lane.}
  \label{fig:imax3_board}
\end{figure*}

\subsection{IMAX Architecture and Compilation Model}

IMAX is a programmable CPU-Grounded Linear Array accelerator.
The evaluated IMAX3 prototype implements one CGLA lane on an AMD Versal Premium VPK180 FPGA with a 64-stage linear chain of PEs and a reported 512\,KiB per-lane LMM capacity at 145\,MHz.

\begin{figure}[!t]
  \centering
  \includegraphics[width=\columnwidth]{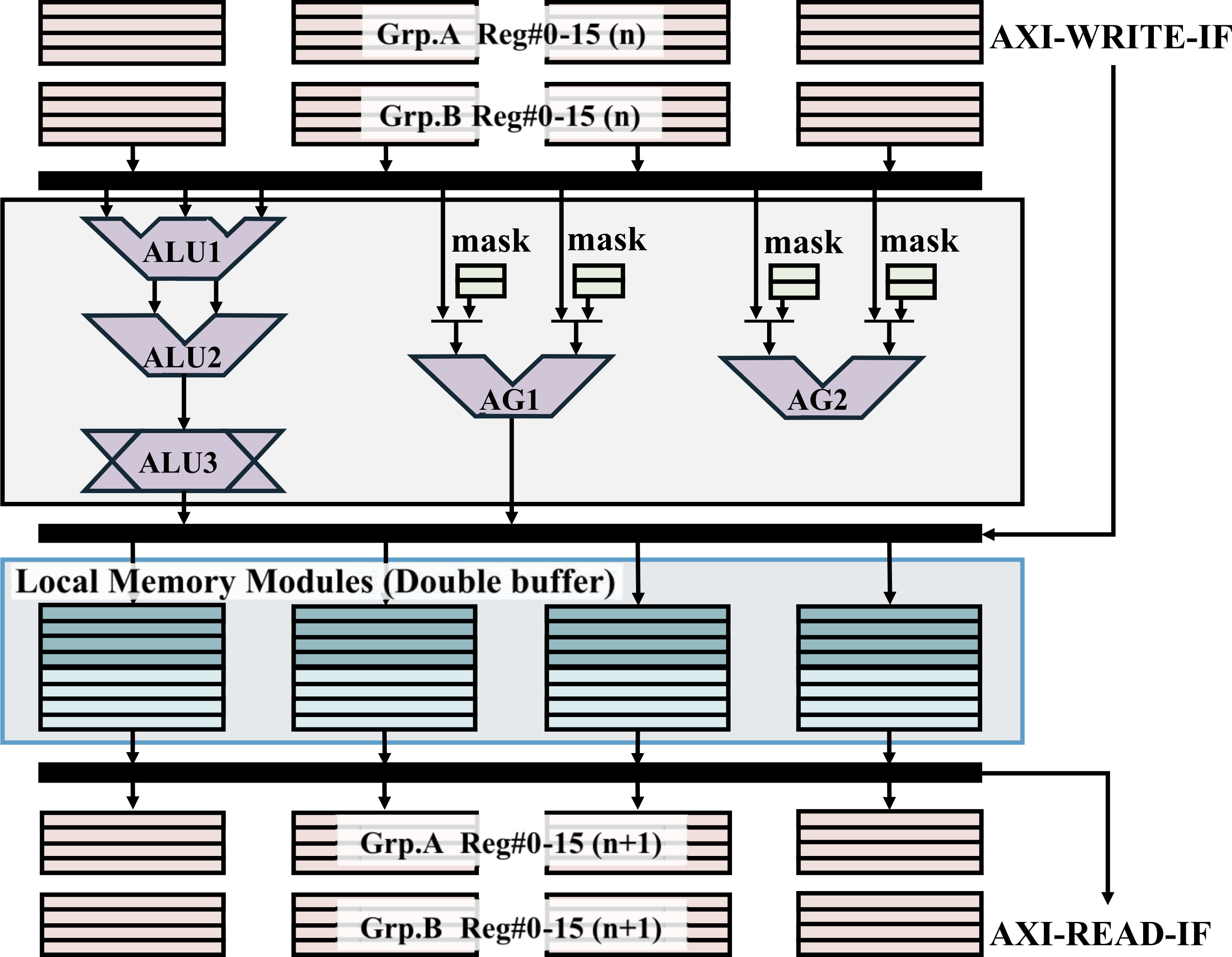}
  \caption{Internal structure of one IMAX compute lane.
           PEs and LMMs are arranged alternately in a one-dimensional array.
           An execution data path connects adjacent PEs and a memory data path connects each PE to its neighboring LMMs.}
  \label{fig:imax_lane}
\end{figure}

Fig.~\ref{fig:imax3_board} shows the IMAX3 prototype, where the host ARM Cortex-A72 in the PS coordinates DMA and kernel launch for the PL array.
Fig.~\ref{fig:imax_lane} shows one compute lane with alternating PEs and LMMs in a one-dimensional array.
The measured kernels execute on one 64-stage spatial pipeline.
Each PE is assigned one static stage by the conv-c2d compiler and acts as an operand-load, arithmetic, or drain/store stage.
The compiler maps the outer row loop, the reduction loop, and the output-column loop to these PE-stage assignments.
Following the execution breakdown used in published IMAX studies, we distinguish configuration, transfer, execution (EXEC), and write-back phases when discussing latency.
Kernels with many small DMA transactions can retain PE-side speedup while showing limited end-to-end gain, so EXEC and DMA-inclusive total times are reported separately.

\subsection{IMAX GEMM Mapping}

The published IMAX GEMM kernel maps $(M_1 \times L) \times (L \times M_2)$ onto the linear pipeline~\cite{imax_access}.
The decode-time projection GEMV used later in this paper corresponds to the special case $M_2=1$.
It first separates the $M_1$ output rows,
then advances the inner-product accumulation along the reduction dimension $L$,
and writes each output value when the accumulation over $L$ completes.
In the projection kernels used in this paper, operand load, FMA-based partial-sum update, and result drain are distributed across a 64-stage one-dimensional pipeline.
When $L$ is long, these stages can remain active for many cycles.
Published IMAX measurements report high pipeline occupancy when $M_1$, $L$, and $M_2$ are all at least 480~\cite{imax_access}.
At $L=64$, the number of partial-sum updates is small, so pipeline fill, drain, and data-movement overhead become relatively large.

\section{Mapping Mamba-2 Kernels onto IMAX}
\label{proposed}

\begin{figure*}[t]
  \centering
  \includegraphics[width=\textwidth]{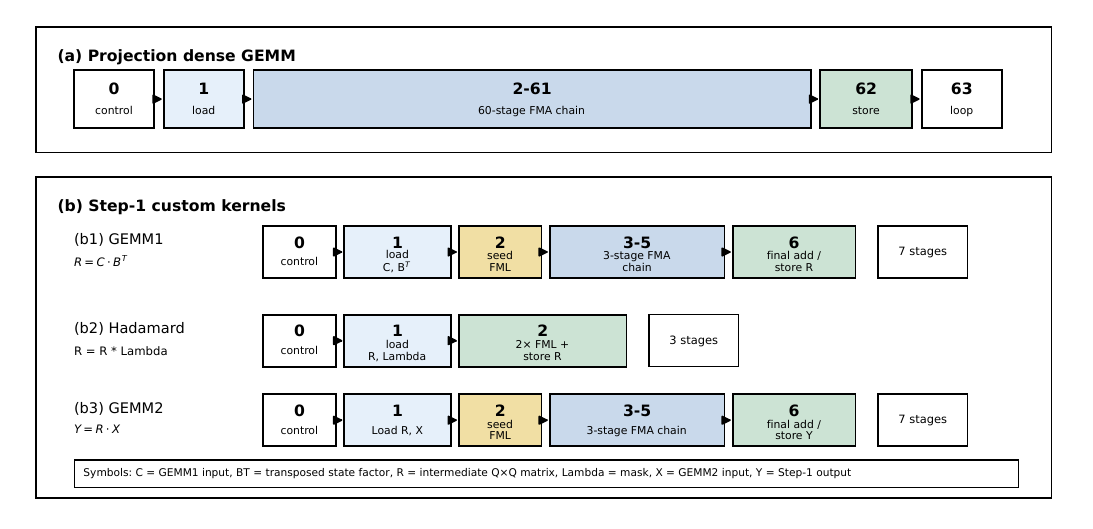}
  \caption{Representative PE-stage view of the generated IMAX kernels used in the projection and Step-1 measurements.
           (Top) Projection occupies stages~0--63 with a 60-stage FMA chain.
           (Bottom) Step-1 uses shallow GEMM1, Hadamard, and GEMM2 kernels.
           Stage~63 is an iteration-control stage in the projection GEMM template, not an SSM recurrence loop.
           Stage numbers are compiler-generated, and physical PE IDs are placement-dependent.}
  \label{fig:pe_stage_map}
\end{figure*}

This section describes the IMAX mappings used for the projection, SSD Step-1, and recurrent-state measurements.

\subsection{Projection GEMM (in-proj and out-proj)}

Both projection phases reuse the standard IMAX GEMM template~\cite{imax_access}.
The outer row dimension is assigned across the available parallel work units.
Accumulation then proceeds along the projection inner dimension.
The output-feature dimension is generated in the final loop.
As summarized in the top panel of Fig.~\ref{fig:pe_stage_map}, the projection path uses stages~0--63.
Stages~2--61 form the 60-stage FMA chain.
Stage~1 loads operands.
Stage~62 performs the final add and store.
Stages~0 and 63 provide iteration control.
The mapping reuses the existing GEMM datapath and changes only compile-time constants such as \texttt{RMGRP}, \texttt{W}, and \texttt{BSTEP}.

\subsection{SSD Step-1 Kernels}

Equation~\eqref{eq:ssd_step1} is mapped to three IMAX kernel calls.
\begin{enumerate}
  \item \textbf{GEMM1}, $R \leftarrow C \cdot B^\top$
    \quad ($Q \times N$) $\times$ ($N \times Q$) $\to$ ($Q \times Q$)
  \item \textbf{Hadamard}, $R \leftarrow R \odot \Lambda$
    \quad (element-wise masking)
  \item \textbf{GEMM2}, $Y \leftarrow R \cdot X$
    \quad ($Q \times Q$) $\times$ ($Q \times d$) $\to$ ($Q \times d$)
\end{enumerate}

The bottom panel of Fig.~\ref{fig:pe_stage_map} uses seven-stage pipelines for GEMM1 and GEMM2 and a three-stage pipeline for Hadamard.
This section defines the mapping.
Section~\ref{ex_and_re} reports the measured boundary cost.

For a single representative head, $M_1=Q=64$ and the reduction length is only 64 in the measured Step-1 sweep.
We therefore increase the outer batch axis to amortize per-launch cost.
The aggregated outer dimension is
\begin{equation}
  M_1 = N_\text{heads} \times Q.
\end{equation}
The evaluation distinguishes the nominal all-head point from the largest sweep point.
The concatenated $C$ matrix remains row-major and follows the standard IMAX GEMM addressing rule.

\label{sec:hadamard_kernel}
The IMAX Hadamard kernel reads $R$ and $\Lambda$ from LMM, computes $R\leftarrow R\odot \Lambda$, and writes $R$ back in-place using two parallel \texttt{OP\_FML} lanes ($Q/2$ pairs per row).
The original two-kernel path adds a DMA round trip because GEMM1 writes $R$ to DRAM, the CPU applies the Hadamard mask, and GEMM2 reads the result again.
Our \texttt{imax\_hadamard()} path keeps this update on IMAX and writes the masked result back in place.
Each call processes one row of the $M_1 \times Q$ result matrix and advances the row pointer in the outer C loop.
This organization avoids the conv-c2d restriction that load and store base addresses remain compile-time constants.
Each iteration processes $Q/2$ element pairs with two parallel \texttt{OP\_FML} lanes, giving 0.25 FLOP/byte arithmetic intensity and removing one DMA round trip per chunk.
The measured boundary overhead still motivates a fused Hadamard--GEMM2 path.
Section~\ref{ex_and_re} reports per-kernel EXEC and wall-clock time for all three IMAX kernels.

\begin{table}[t]
  \centering
  \caption{Per-layer FLOP distribution, Mamba-370M ($d_\text{model}=1{,}024$, $d_\text{inner}=2{,}048$, $H=32$, $d_\text{head}=64$, $N=128$, $Q=64$).}
  \label{tab:flop}
  \small
  \renewcommand{\arraystretch}{1.15}
  \begin{tabular}{lrr}
    \toprule
    Phase & FLOP expression & $T{=}512$ (MF) \\
    \midrule
    in-proj          & $2T d_\text{model} d_\text{in}$  & 4{,}598 \\
    out-proj         & $2T d_\text{inner} d_\text{model}$  & 2{,}147 \\
    SSD Step-1 GEMM1 & $2T H Q N$              &   268 \\
    SSD Step-1 GEMM2 & $2T H Q d$              &   134 \\
    Step-3 (state)   & $2 T N d_\text{head} H$ &   268 \\
    \midrule
    \textbf{Total}   &                       & \textbf{7{,}415} \\
    \rowcolor{black!8}
    \textbf{Proj.\ share} &                  & \textbf{91\%} \\
    \bottomrule
  \end{tabular}
\end{table}

\begin{table*}[!t]
  \centering
  \caption{Evaluation platforms and ASIC model context.}
  \label{tab:processor_comparison_annotated}
  \footnotesize
  \renewcommand{\arraystretch}{1.15}
  \setlength{\tabcolsep}{4.2pt}
  \begin{adjustbox}{width=\textwidth}
  \begin{tabular}{@{} l l l r r l l @{}}
    \toprule
    \textbf{Device} & \textbf{Host / CPU} & \textbf{Role} & \textbf{Process} & \textbf{Clock setting} & \textbf{Memory} & \textbf{Measurement scope} \\
    & & & \textbf{(nm)} & \textbf{(MHz or mode)} & & \\
    \midrule
    \textbf{ARM Cortex-A72 (on Versal)} & Cortex-A72 & CPU reference & 7 & 1400 & DDR4 & Measured software time \\
    \textbf{IMAX FPGA prototype} & Cortex-A72 & Kernel and dispatch path & 7 & 145 & 8\,GB + 4\,GB DDR4 & Measured FPGA time \\
    \rowcolor{black!8}
    \textbf{IMAX ASIC model} & -- & Context only & 28 & 840 & -- & Design Compiler core estimate \\
    \bottomrule
  \end{tabular}
  \end{adjustbox}
\end{table*}

\subsection{Step-3 Recurrent State Update}

The inter-chunk recurrence~\eqref{eq:step3} is a small $N \times d$ matrix-vector product per chunk with sequential dependency, so transfer scheduling rather than peak compute determines its cost.
For Mamba-370M, the per-head shape is only $128 \times 64$, which is too small to saturate the 64-stage FMA pipeline.
For the layer-level composition, Step-3 stays on the host CPU.
Section~\ref{sec:ssm_state} reports IMAX recurrent-state runs as a separate transfer-scheduling study.
If more of this phase is moved to IMAX, double-buffered LMM use and ping-pong scheduling are the main next directions.
During decode at $T=1$, projection GEMV becomes less compute dense, but transfer overhead still dominates the current FPGA implementation.

\section{Experiments}
\label{ex_and_re}

\subsection{Experimental Setup}
Table~\ref{tab:flop} reports nominal Mamba-370M FLOP accounting with $Q{=}64$.
Table~\ref{tab:processor_comparison_annotated} lists the platform scope used in the evaluation.
The ARM Cortex-A72 software path is the correctness and CPU-timing reference.
IMAX latency and correctness are measured on the FPGA prototype.
The ASIC context in Table~\ref{tab:processor_comparison_annotated} comes from synthesis of the CGLA core with Synopsys Design Compiler and a 28\,nm standard-cell flow, with a reported maximum clock of 840\,MHz~\cite{imax_access,SynopsysNDDCUltra}.
It is not a silicon or board-power measurement and excludes the host CPU, DDR, DMA, and board components.
Consequently, this paper does not use that core estimate for cross-platform system-energy claims.
All kernels use FP32.
The projection values are means of three archived runs.
Their DMA-inclusive min--max spread is below 0.1\% for both projection kernels.
The Step-1 and token-generation values are representative runs, and no confidence interval is claimed for them.
Outputs are checked against the ARM Cortex-A72 software reference.
The recurrent-state measurements report end-to-end IMAX time because that kernel performs inline DMA.
The layer composition uses a pipeline-aligned timing setting, $T{=}120$ and $Q{=}60$.
Mamba-370M is used for layer-level kernel characterization, and Mamba-130M is used for the token-generation run.

\subsection{IMAX Kernel Measurements}
\textbf{Projection GEMM.}
Projection GEMM latency is measured using the IMAX GEMM kernel~\cite{imax_access}.
Dimensions are rounded to pipeline-aligned values ($L$ to a multiple of $60$ and $M_2$ to a multiple of $8$).
For in-proj, the IMAX run records 37.8\,ms EXEC and 1.17\,s DMA-inclusive total time, so DMA dominates the measured total time rather than arithmetic.
The archived repeated projection runs have a min--max spread below 0.1\% in DMA-inclusive total time for both Mamba-370M projection kernels.
Projection is the long-reduction CGLA fit in this workload, while DMA remains the current in-proj cost.

\textbf{SSD Step-1.}
We sweep $M_1$ from 64 to 4{,}096 with the representative Step-1 kernel setting $N=Q=d_\text{head}=64$.
The unfused three-kernel Step-1 path is not end-to-end beneficial.
GEMM1, Hadamard, and GEMM2 record 7{,}068, 1{,}178, and 7{,}052\,\si{\micro\second} EXEC, respectively ($Q{=}N{=}d_\text{head}{=}64$, batch size $16$, FP32), but boundary transfers raise the aggregate wall-clock time to 48{,}922\,\si{\micro\second}.
The matched CPU reference in this rerun takes 19{,}407\,\si{\micro\second}.
In the separate single-GEMM sweep, the EXEC component achieves a 1.43$\times$ speedup over the ARM Cortex-A72 reference at the largest sweep point ($M_1{=}4{,}096$), but this component-only result does not overcome boundary and transfer overheads.
The limiting factors are short reductions and kernel boundaries.
Fusing Hadamard with GEMM2 is the next optimization target.

\label{sec:ssm_state}
Representative recurrent-state runs use $Q{=}64$, $N{=}128$, and $d_\text{head}{=}64$ with inline DMA.
Block autotuning changes wall-clock time from 29.3 to 28.1\,ms for Mamba-370M and from 124.4 to 107.0\,ms for Mamba-1.4B relative to the default block setting.
The resulting speedup range is 1.04--1.16$\times$, and $\max|\Delta| = 3.46\times 10^{-3}$ between IMAX and CPU reference outputs.
The 1.04$\times$ Mamba-370M change should be treated as no measurable gain without run-to-run variance support.
The 1.16$\times$ Mamba-1.4B trend identifies Step-3 as a transfer-scheduling target rather than a compute-bound one.

\subsection{Phase Bottlenecks and Scope}
\label{sec:layer_composition}
At the pipeline-aligned timing setting $T{=}120$, $Q{=}60$, projection remains the dominant FLOP component and dominates IMAX EXEC time.
SSD Step-1 remains a smaller FLOP component and is limited by short reductions and kernel boundaries.
The comparison is therefore made against the co-located ARM reference from the same kernel programs rather than against a GPU system with a different power boundary.
Because only FP32 is evaluated, the results do not establish the behavior of FP16, BF16, or quantized edge-inference paths.

\subsection{Token-Level Integration with IMAX Offload}
\label{sec:e2e_token_generation}
\begin{figure}[!t]
  \centering
  \includegraphics[width=0.84\columnwidth]{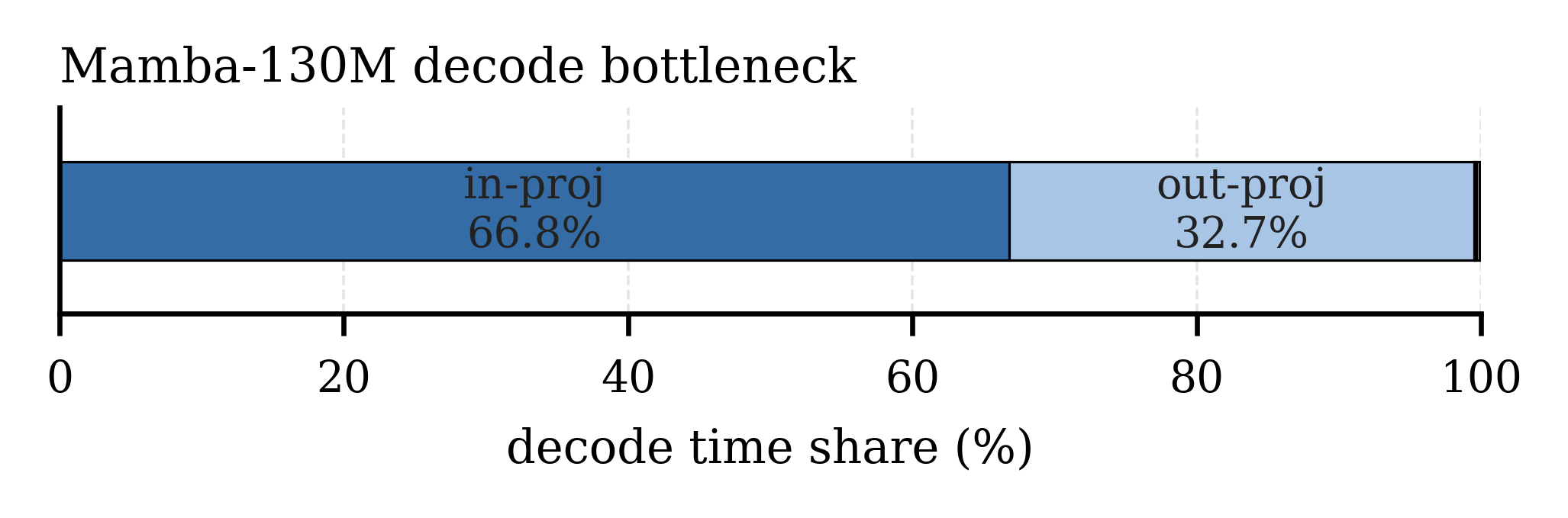}
  \caption{Mamba-130M decode-stage time share for the IMAX hardware-dispatch path.
           In-proj and out-proj account for 66.8\% and 32.7\% of the time, while each remaining segment is below 0.3\%.}
  \label{fig:mamba130m_decode_breakdown}
\end{figure}

We place the IMAX offload path inside a standalone C Mamba-130M autoregressive decoding program on the IMAX FPGA platform.
The program loads the model weights and tokenizer, runs greedy decoding, and dispatches the profiled kernels to IMAX.
The run preserves the ARM Cortex-A72 software-reference greedy token sequence and reports zero output errors.
Throughput reaches 0.345 tokens/s for the 200-token generation run, and Fig.~\ref{fig:mamba130m_decode_breakdown} identifies projection GEMV as the bottleneck.
This integration run has no matched end-to-end CPU or GPU throughput measurement under the same program and run conditions, so it is not used to claim system-level acceleration.

In-proj and out-proj account for 66.8\% and 32.7\% of decode time, while x-proj, dt-proj, state update, and convolution are each below 0.3\%.
Future work will evaluate persistent transposed weights, wider row grouping, Step-1 kernel fusion, and reduced-precision execution under matched end-to-end ARM Cortex-A72 and Jetson AGX Orin baselines.
These directions target projection GEMV and transfer boundaries rather than the SSM state update itself.

\section{Conclusion}
\label{conclusion}

This paper measured Mamba-family SSM kernels on the IMAX programmable CGLA from kernel execution to token generation.
The results separate long-reduction projections that fit the pipeline, unfused SSD Step-1 limited by short reductions and kernel boundaries, and autoregressive decoding dominated by projection GEMV.
The single-lane Mamba-130M run reaches 0.345 tokens/s while preserving the ARM reference greedy token sequence.
The unfused Step-1 path is slower than its ARM reference after boundary transfers, and the token run has no matched end-to-end throughput baseline.
DMA-inclusive execution and kernel-boundary overheads limit the FPGA prototype, so this is a suitability and bottleneck study rather than a demonstrated system-level acceleration.
The FP32-only evaluation and core-only ASIC model do not support reduced-precision or board-energy conclusions.
Future work will evaluate persistent projection weights, wider row grouping, Step-1 fusion, and reduced precision with matched system-level baselines.

\section*{Acknowledgment}
This work was supported by the JST-ALCA-Next Program (Grant Number JPMJAN23F4) and JSPS KAKENHI (Grant No.\ 22H00515).
We also acknowledge the activities of VDEC, The University of Tokyo, in collaboration with NIHON SYNOPSYS G.K.

\bibliographystyle{IEEEtran}
\bibliography{bibliography}

\begin{thebibliography}{10}
\providecommand{\url}[1]{#1}
\csname url@samestyle\endcsname
\providecommand{\newblock}{\relax}
\providecommand{\bibinfo}[2]{#2}
\providecommand{\BIBentrySTDinterwordspacing}{\spaceskip=0pt\relax}
\providecommand{\BIBentryALTinterwordstretchfactor}{4}
\providecommand{\BIBentryALTinterwordspacing}{\spaceskip=\fontdimen2\font plus
\BIBentryALTinterwordstretchfactor\fontdimen3\font minus
  \fontdimen4\font\relax}
\providecommand{\BIBforeignlanguage}[2]{{%
\expandafter\ifx\csname l@#1\endcsname\relax
\typeout{** WARNING: IEEEtran.bst: No hyphenation pattern has been}%
\typeout{** loaded for the language `#1'. Using the pattern for}%
\typeout{** the default language instead.}%
\else
\language=\csname l@#1\endcsname
\fi
#2}}
\providecommand{\BIBdecl}{\relax}
\BIBdecl

\bibitem{mamba1}
A.~Gu and T.~Dao, ``{Mamba}: Linear-time sequence modeling with selective state
  spaces,'' in \emph{Proc. Conference on Language Modeling (COLM)}, 2024.

\bibitem{mamba2}
T.~Dao and A.~Gu, ``Transformers are {SSMs}: Generalized models and efficient
  algorithms through structured state space duality,'' in \emph{Proc. 41st Int.
  Conf. Machine Learning (ICML)}, ser. Proc. Machine Learning Research, vol.
  235, 2024, pp. 10\,041--10\,071.

\bibitem{imax_access}
T.~Akabe, V.~{Trung Duong LE}, and Y.~Nakashima, ``{IMAX}: A power-efficient
  multilevel pipelined {CGLA} and applications,'' \emph{IEEE Access}, vol.~13,
  pp. 31\,899--31\,911, 2025.

\bibitem{lightmamba}
R.~Wei, S.~Xu, L.~Zhong, Z.~Yang, Q.~Guo, Y.~Wang, R.~Wang, and M.~Li,
  ``{LightMamba}: Efficient {Mamba} acceleration on {FPGA} with quantization
  and hardware co-design,'' in \emph{Proc. Design, Automation and Test in
  Europe (DATE)}, 2025.

\bibitem{fastmamba}
A.~Wang, H.~Shao, S.~Ma, and Z.~Wang, ``{FastMamba}: A high-speed and efficient
  {Mamba} accelerator on {FPGA} with accurate quantization,'' in \emph{Proc.
  IEEE Computer Society Annual Symposium on VLSI (ISVLSI)}, 2025.

\bibitem{marca}
J.~Li, S.~Huang, J.~Xu, J.~Liu, L.~Ding, N.~Xu, and G.~Dai, ``{MARCA}: {Mamba}
  accelerator with reconfigurable architecture,'' in \emph{Proc. IEEE/ACM Int.
  Conf. Computer-Aided Design (ICCAD)}, 2024, pp. 1--9, doi:
  10.1145/3676536.3676798.

\bibitem{marcav2}
J.~Li, S.~Huang, J.~Xu, J.~Liu, N.~Xu, and G.~Dai, ``{MARCA-v2}: {Mamba}
  accelerator with complementary state space model sparsity and reconfigurable
  architecture,'' \emph{IEEE Trans. Comput.-Aided Design Integr. Circuits
  Syst.}, 2025, {Early Access}, doi: 10.1109/TCAD.2025.3624278.

\bibitem{mambaopu}
S.~Lu, X.~Yu, T.~Zhao, S.~Miao, X.~Sheng, C.~Wu, L.~Zhao, T.-J. Lin, and L.~He,
  ``{MambaOPU}: An {FPGA} overlay processor for state-space-duality-based
  {Mamba} models,'' in \emph{Proc. ACM/IEEE Design Automation Conference
  (DAC)}, 2025, pp. 1--7.

\bibitem{specmamba}
L.~Zhong, S.~Xu, H.~Wen, T.~Xie, Q.~Guo, Y.~Wang, and M.~Li, ``{SpecMamba}:
  Accelerating {Mamba} inference on {FPGA} with speculative decoding,'' in
  \emph{Proc. IEEE/ACM International Conference on Computer-Aided Design
  (ICCAD)}, 2025, pp. 1--9.

\bibitem{epochcore}
S.~Raja, C.~Demirkiran, A.~Sarkar, M.~Popovic, and A.~Joshi, ``Systolic
  array-based accelerator for structured state-space models,'' 2025,
  arXiv:2507.21394.

\bibitem{ieeeaccess}
T.~Ando, Y.~Eto, A.~Takeuchi, and Y.~Nakashima, ``Efficient kernel mapping and
  comprehensive system evaluation of {LLM} acceleration on a {CGLA},''
  \emph{IEEE Access}, vol.~13, pp. 199\,631--199\,646, 2025.

\bibitem{whisper}
T.~Ando, Y.~Eto, A.~Takeuchi, and Y.~Nakashima, ``Energy-efficient hardware
  acceleration of {Whisper} {ASR} on a {CGLA},'' in \emph{Proc. 13th Int. Symp.
  Computing and Networking (CANDAR)}, 2025, pp. 85--91.

\bibitem{TPU}
N.~P. Jouppi, G.~Kurian, S.~Li, P.~Ma, R.~Nagarajan, L.~Nai, N.~Patil,
  S.~Subramanian, A.~Swing, B.~Towles, C.~Young, X.~Zhou, Z.~Zhou, and
  D.~Patterson, ``{TPU v4}: An optically reconfigurable supercomputer for
  machine learning with hardware support for embeddings,'' in \emph{Proc. 50th
  Annu. Int. Symp. Computer Architecture (ISCA)}, 2023, pp. 1147--1160, doi:
  10.1145/3579371.3589350.

\bibitem{diannao}
T.~Chen, Z.~Du, N.~Sun, J.~Wang, C.~Wu, Y.~Chen, and O.~Temam, ``{DianNao}: {A}
  {S}mall-{F}ootprint {H}igh-{T}hroughput {A}ccelerator for {U}biquitous
  {M}achine-{L}earning,'' in \emph{Proceedings of the 19th International
  Conference on Architectural Support for Programming Languages and Operating
  Systems (ASPLOS)}, 2014, pp. 269--284.

\bibitem{eyeriss}
Y.-H. Chen, J.~Emer, and V.~Sze, ``Eyeriss: A spatial architecture for
  energy-efficient dataflow for convolutional neural networks,'' in
  \emph{Proceedings of the 43rd Annual International Symposium on Computer
  Architecture}, 2016, pp. 367--379.

\bibitem{cgra_survey}
A.~Podobas, K.~Sano, and S.~Matsuoka, ``A survey on coarse-grained
  reconfigurable architectures from a performance perspective,'' \emph{IEEE
  Access}, vol.~8, pp. 146\,719--146\,743, 2020.

\bibitem{cgra1}
C.~Torng, P.~Pan, Y.~Ou, C.~Tan, and C.~Batten, ``Ultra-elastic {CGRAs} for
  irregular loop specialization,'' in \emph{Proc. IEEE Int. Symp. High-Perform.
  Comput. Archit. (HPCA)}, 2021, pp. 412--425.

\bibitem{cgra_cnn2}
J.~Lee and J.~Lee, ``Specializing {CGRAs} for light-weight convolutional neural
  networks,'' \emph{IEEE Trans. Comput.-Aided Design Integr. Circuits Syst.},
  vol.~41, no.~10, pp. 3387--3399, 2022.

\bibitem{finegrainedfusion}
R.~Geens, A.~Symons, and M.~Verhelst, ``Fine-grained fusion: The missing piece
  in area-efficient state space model acceleration,'' in \emph{Proc. 34th
  International Conference on Parallel Architectures and Compilation Techniques
  (PACT)}, 2025, pp. 281--291, doi: 10.1109/PACT65351.2025.00034.

\bibitem{SynopsysNDDCUltra}
{Synopsys, Inc.}, ``{Design Compiler}: {Concurrent Timing, Area, Power, and
  Test Optimization},'' [Online]. Available:
  \url{https://www.synopsys.com/implementation-and-signoff/rtl-synthesis-test/dc-ultra.html},
  accessed: May 25, 2025.

\end{thebibliography}

\end{document}